# Quantitative Phase Microscopy Spatial Signatures of Cancer Cells

Darina Roitshtain[1], Lauren Wolbromsky[1], Evgeny Bal[1],
Hayit Greenspan[1], Lisa L. Satterwhite[2], Natan T. Shaked[1,*]

[1] Department of Biomedical Engineering, Faculty of Engineering, Tel Aviv University, Tel Aviv 69978, Israel

[2] Department of Biomedical Engineering, Duke University, Durham, NC 27708, USA

[*]nshaked@tau.ac.il

## Abstract

We present cytometric classification of live healthy and cancer cells by using the spatial morphological and textural information found in the label-free quantitative phase images of the cells. We compare both healthy cells to primary tumor cell and primary tumor cells to metastatic cancer cells, where tumor biopsies and normal tissues were isolated from the same individuals. To mimic analysis of liquid biopsies by flow cytometry, the cells were imaged while unattached to the substrate. We used low-coherence off-axis interferometric phase microscopy setup, which allows a single-exposure acquisition mode, and thus is suitable for quantitative imaging of dynamic cells during flow. After acquisition, the optical path delay maps of the cells were extracted, and used to calculate 15 parameters derived from cellular 3-D morphology and texture. Upon analyzing tens of cells in each group, we found high statistical significance in the difference between the groups in most of the parameters calculated, with the same trends for all statistically significant parameters. Furthermore, a specially designed machine learning algorithm, implemented on the phase map extracted features, classified the correct cell type (healthy/cancer/metastatic) with 81%-93% sensitivity and 81%-99% specificity. The quantitative phase imaging approach for liquid biopsies presented in this paper could be the basis for advanced techniques of staging freshly isolated live cancer cells in imaging flow cytometers.

## Introduction

Finding cancer in its early curable stages is a clear and critical unmet need. Late stage metastatic forms of cancer are almost always uniformly fatal (1,2). Many cancers are found in early stages only due to incidental or proactive screening, especially in high risk groups (3,4). However, some cancers including pancreatic cancer and colon cancer can develop asymptomatically and thus often escape early detection (5,6). The primary basis for diagnosis of cancer is evaluation of morphological changes in a tissue biopsy by a trained pathologist, a process with inherent subjectivity, performed usually after the location of the suspect tissue is known (7).

Flow cytometry of body fluids obtained by routine medical tests can identify circulating tumor cells after their separation from the other fluid contents (8,9). Interest in using circulating tumor cells found in liquid biopsies to diagnose solid tumors has exploded recently, especially to address difficulties in analyzing tumor biopsies due to intractable anatomical placement or to determine if metastatic disease is present (10–13). In 2004, a landmark study found that the number of circulating tumor cells in blood could predict survival in metastatic breast cancer (14). Cancer-specific signatures in liquid biopsies include DNA sequence, composition of cancer cell exosomes, and unique systemic response reflected by components in blood, tears, saliva or urine assayed by genomics, epigenetics, proteomics and metabolomics (4,15). Increasingly, liquid biopsy has been suggested as a viable and affordable mechanism to safeguard good health and to detect cancer in its asymptomatic early curable stages through proactive monitoring, especially in high risk groups.

Detection of circulating tumor cells requires a highly sensitive method to identify a small number of diseased cells in a large cell population. Isolation of these cancer cells is laborious and typically yields uniformly round cells, which are hard to stage without advanced methods. Thus, other internal features of the cells should be used for evaluation (4).

During the progression of a healthy cell to immortal cancerous cell and later to metastatic cell, the biophysical and morphological phenotypes of the cell change (16). Many scientific efforts have been made to perform cytometry of cancer cells with the goal of revealing the unique biophysical properties of these cells for cancer prognosis include measuring the mechanical (17–20) or optical properties of the cells by label-based (22) or label-free (23–29) techniques. For example, multiphoton laser tomography combined with fluorescence lifetime imaging has been used to generate 30 optical parameters that are able to distinguish normal nevi from melanoma *in situ* with high sensitivity and specificity (30). Atypical cells, underlying cytoskeletal disorganization in the epidermis, poorly defined keratinocyte cell borders and presence of dendritic cells typified melanoma tissue compared to normal nevi and resulted in the unique diagnostic optical parameters. Additionally, basal cell carcinoma could be distinguished from melanoma using these parameters. Although markers provide biological assays with a high degree of specificity, using fluorescent markers might cause cytotoxicity by perturbing the cell environment, by influencing its behavior over time and its viability, and eventually damaging the accuracy of the test or prohibiting further clinical use of the isolated cells (31). In flow cytometry for cell sorting, one evaluates cellular features through fluorescence markers and purifies the heterogeneous cell suspension into fractions containing a single cell type (32). However, in addition to possible cytotoxicity, suitable markers might be not available for certain cell types, and some markers might be difficult to use (33). Specifically, fluorescent markers tend to photobleach, which damages the image contrast and the prognosis results (34).

The internal morphology and texture of cancer cells changes during oncogenesis (35-37). Specifically, the intrinsic refractive index of live cells can indicate abnormal cell morphology. The cell refractive index is related to the optical interaction of the light field with cellular organelles and chemical composition, and thus can be potentially used for quantitative monitoring and diagnosis of the cellular phenotypes (28) and indicate abnormal cell morphology. In addition, dry mass of the cancer cells has been recognized as a possible diagnostic and monitoring marker (38,39).

These cellular changes in cancer cells can be potentially detected by label-free imaging techniques. Without staining, however, biological cells are nearly transparent under bright-field microscopy, as their absorption differs only slightly from that of their surroundings, resulting in a low image contrast. An internal contrast mechanism that can be used when imaging cells without staining is their refractive index. The light beam passing through the imaged cells is delayed, since the cells have a slightly higher refractive index compared with their surroundings. Conventional intensity-based detectors are not fast enough to record this delay directly. Phase imaging methods, on the other hand, use optical interference to record the delays of light passing through the sample, and thus they are able to create label-free contrast in the image (40). In contrast to qualitative phase contrast methods, such as Zernike's phase contrast and differential interference contrast (DIC) microscopies, quantitative phase microscopy yields the optical thickness map or optical path delay (OPD) map of the cell on all spatial (x-y) points. Per each x-y point of this map, OPD is equal to the integral of the refractive-index values across the cell thickness.

Quantitative phase imaging techniques, interferometric based (23,28,41–43) and non-interferometric based (24), (29) have been used to analyze cell features for red blood cells (41), stem cells (42), cancer tissues (23) and cancer cells (24,28,29,43), and showed the ability to differentiate between various conditions of cells and tissues, based on the average OPD values (23,42,43), or other parameters that are based on the OPD map and the cell visible morphology (24,29), as well as their spatial-frequency content (41).

Machine learning of cytometric data from circulating tumor cells enables automatic analysis of a large number of cells with good classification results (44). Specific to interferometry, after extraction of various cellular features from the cellular OPD maps, machine learning techniques can assign weights to these features for classification. Several machine learning techniques have been previously applied on reconstructed digital interferograms starting from 2005 to identify filamentous microorganisms (45); and later to classify stained and unstained cells, and quantify cell viability and concentration (46); and to grade red blood cells infected by the malaria parasite (47).

In our study, we compared the quantitative phase imaging-based features of healthy and cancer cells and of primary cancer and metastatic cancer cells. When doing this comparison, we have chosen pairs of cell lines taken from the same individual to avoid differences that are related to changes between people's organs and disease expression. In order to obtain data for significant statistics, cell imaging techniques should have high throughput capabilities, but still be affordable (16). Acquiring the cells during fast flow can enable high throughput. In our research, the cells were alive and not attached to the surface, which allowed for analyzing a large number of cells during flow, in contrast to previous studies that used adherent cells (28,43), fixed cells (24,29), or tissue samples (23), which were limited in the amount of recorded data. In addition, we have used off-axis interferometry, which requires only a single camera exposure, and thus is suitable for acquisition of rapid dynamics, such as the one occurring during cell flowing.

After acquiring off-axis interferograms of the cells and extracting their OPD maps, we calculated cellular features based on these quantitative maps, and applied machine learning approaches for cell group classification. Our quantitative phase imaging approach is expected to yield an automated tool to distinguish oncogenic progression and metastasis based on label-free cancer cell cytometry.

## Materials and methods

Low-coherence interferometric phase microscopy (IPM) setup

We used low-coherence interferometric phase microscopy (IPM) to capture the OPD map of healthy, primary cancer and metastatic cells *in vitro*. The (IPM) system used in the experiments is a low-coherence Mach-Zehnder imaging interferometer for creating off-axis image interferograms of the sample. Figure 1 presents a scheme of a system, which is illuminated by a supercontinuum laser source, coupled to an acousto optical tunable filter (AOTF) (Fianium SC-400-4 and AOTF, 650 nm with FWHM spectral bandwidth of 7nm). The beam is split into a reference and a sample beam by BS1. The sample beam passes through the cell sample S, which is located on an XYZ micrometer stage. This beam is then magnified by microscope objective MO1 (Newport, 60×, 0.85 NA). In parallel, the reference beam propagates through identical microscope objective MO2. Both beams are projected through tube lens L (f=150 mm) onto a CCD digital camera (Thorlabs, DCC1545M, max framerate – 25 fps, exposure time – 40 ms). The lateral resolution was 0.76 µm. Since low-coherence illumination is used to minimize spatial noise and parasitic interferences, retroreflectors RRs on micrometers are utilized to adjust the beam paths of the sample and reference beams. These two beams interfere on the camera at a small off-axis angle and induce straight off-axis fringes. The temporal and spatial OPD (axial) stabilities were 0.62 nm and 1.1 nm, respectively.

Cell culture

We measured three pairs of isogenic cell lines: normal skin cells Hs 895.Sk (ATCC CRL-7636) and melanoma skin cells Hs 895.T (ATCC CRL-7637), melanoma skin cells WM-115 (ATCC CRL-1675) and metastatic melanoma skin cells WM-266-4 (ATCCCRL-1676), colorectal adenocarcinoma colon cells SW-480 (ATCC CCL-228) and metastatic from lymph node of colorectal adenocarcinoma cells SW-620 (ATCC CCL-227). The first pair of cells was chosen to compare healthy cells versus cancer cells, and the other two pairs of cells were chosen to compare primary cancer cells versus metastatic cells. It is important to note that each of the cell line comparisons is from the same individual.

The complete growth medium used for the Hs cell pair is Dulbecco's Modified Eagle's Medium (DMEM) (ATCC, SN. 30-2002) supplemented with 10% Fetal Bovine Serum (FBS) (BI, SN. 04-007-1A) .

The complete growth medium used for the WM cell pair is DMEM (DMEM) (BI, SN. 01-055-1A) supplemented with 10% FBS (BI, SN. 04-007-1A) and 2 mM L-glutamine (BI, SN. 03-020-1B) (48,49).

The complete growth medium used for the SW cell pair is BI Roswell Park Memorial Institute (RPMI) 1640 Medium without L-glutamine (BI, SN. 01-104-1A) supplemented with 10% FBS (BI, SN. 04-007-1A) and BI 2 mM L-glutamine (BI, SN. 03-020-1B) (50–55).

The cell lines were incubated under standard cell culture conditions at 37°C and 5% $CO_2$ in a humidified incubator until 80% confluence was achieved.

Preparation for imaging

Prior to the imaging experiment, the cells were trypsinized for suspension, supplemented with a suitable medium, and inserted into an adhesive chamber (Grace Bio-Labs HybriWell™ sealing system, SecureSeal™ adhesive chamber, chamber volume 18 µL, 13 mm diameter × 0.15 mm thickness, ports diameter 1.5 mm, Sigma Aldrich SN. GBL611101) attached to a cover slip. This chamber induced a constant thickness value on the entire imaged sample, which is important for the flatness of the final phase map. Another adhesive chamber was filled with the suitable medium and placed in the reference beam propagation path. Then, all cells lines were quantitatively imaged without labeling using the low-coherence IPM system shown in Fig. 1.

Data analysis

To extract the quantitative phase map from the acquired off-axis image interferograms, we used the off-axis interferometry Fourier-based algorithm (56), which includes a 2-D Fourier transform, filtering one of the cross-correlation terms, and an inverse 2-D Fourier transform, where the argument of the resulting matrix is the wrapped phase of the sample. To compensate for stationary aberrations and field curvatures, we subtracted from the wrapped phase map of the sample, a phase map which is extracted from an interferogram acquired with no sample. We then applied the unweighted least squares phase unwrapping algorithm to resolve 2π phase ambiguities (57). The resulting unwrapped phase map is multiplied by the wavelength and divided by 2π to obtain the quantitative OPD map of the sample, which is defined as follows:

$$(1)\ OPD_c(x,y) = [\overline{n}_c(x,y) - n_m] \times h_c(x,y),$$

where $n_m$ is the refractive index of the medium, $h_c$ is the thickness profile of the cell, and $\overline{n}_c$ is the cell integral refractive index, which is defined as follows (58):

$$(2)\quad \overline{n}_c(x,y) = \frac{1}{h_c}\int_o^h n_c(x,y,z)dz.$$

To separate single cells from the background and be able to process only the OPD related to the cells, we used the Normalized Cut Algorithm as an edge detector (59). This method is based on a graph formulation, wherein the nodes of the graph are the points in the feature space with a similarity function weight connecting them. The goal is to partition the vertices into disjoint sets $V_1, V_2, ..., V_m$, whereby the similarity within a set $V_i$ is high, and across different sets is low. After implementation of this algorithm, a morphological opening operator was applied in order to connect the gradient lines that were detected. Next, a morphological dilation operator was applied in order to expand the connected lines. At last, a global threshold was applied in order to remove background pixels that erroneously appeared in the cell area.

Using the above described methods, we could create a data set containing the OPD information of the cell areas only, and calculate the following parameters that are based directly on the OPD map defined in Eq. (1), without decoupling the cellular thickness profile from the refractive index as a prior stage:

1. **Mean** and **median** of $OPD_c$**.**
2. **Dry mass**: This parameter quantifies the mass of the non-aqueous material of the cell, yielding information about cell growth (60,61).

   $$(3)\ M = \frac{1}{\alpha}\int_{S_c} OPD_c(x,y)ds = \frac{S_c}{\alpha} \times < OPD_c >,$$

   where $\alpha$ is the refractive increment and approximated as $0.18\text{-}0.21\ \text{ml/g}$, $S_c$ is the **projected cell area** on the x-y plane, and $< OPD_c >$ is the averaged OPD over the cell area. We used $\alpha = 0.2$ .
3. **Dry mass averaged density:** This parameter can be calculated using the cell dry mass as follows (62):

   $$(4)\ \overline{\sigma}_M = \frac{M}{S_c}.$$
4. **Phase volume**: This is not the actual cell volume but only the equivalent of the cell volume that is based on $OPD$ directly and takes into consideration refractive-index variations inside the cell (62).

   $$(5)\ V_\varphi = \int_{S_c} OPD_c(x,y)ds.$$

As can be seen, the phase volume is proportional to the dry mass of the cell.

5. **Phase surface area**: This parameter can be calculated as the sum of the upper surface area of the phase profile and the projected area, as follows:

$$(6)\ SA_{\varphi} = \int_{S_c} dA + S_c = \int_{S_c} (1 + \delta h_x^2 + \delta h_y^2)^{1/2} dxdy + S_c \ ,$$

where $dA$ is the discrete cell surface area as projected over a single camera pixel, $\delta h_x$ and $\delta h_y$ are the gradients along the $x$ and $y$ directions of the cell OPD map (63,64).

6. **Phase surface area to volume ratio**: This parameter is a generalized version of the physical surface area to volume parameter (65,66), but again it takes into consideration phase changes in the cell:

$$(7)\ SAV = \frac{SA_{\varphi}}{V_{\varphi}}.$$

7. **Phase surface area to dry mass ratio**: This parameter is defined as follows (62):

$$(8)\ SDM = \frac{SA_{\varphi}}{M}.$$

The last two parameters can quantify cell metabolism and describe how much material a surface unit transfers to one volume unit or mass unit.

8. **Projected area to volume ratio**: This parameter describes the flatness of the cell (62).

$$(9)\ PAV = \frac{S_c}{V_{\varphi}}.$$

9. **Phase sphericity index**: This parameter quantifies the degree of cell roundness. The sphericity of an object is the ratio of object volume and the surface area. Round shape is a value that may imply on cell abnormality (62,67–69). It is a dimensionless constant with values ranging from zero for a laminar disk to unity for a sphere (68).

$$(10)\ \psi = \pi^{1/3} \times \frac{(6 \times V_{\varphi})^{2/3}}{SA_{\varphi}}.$$

10. **Phase statistical parameters**: These parameters describe the dry mass or volume distribution in the cell. They are based on changes in phase values and thus react to structural alternations of cell organelles and factors. To use these parameters, the phase values over the projected cell area need to be written as a single vector containing $n$ values. Then, the following statistical parameters can be defined (62):
    a. **Phase variance**: This parameter measures how a set of the cell OPD values is spread out.

    $$(11)\ \sigma_{\varphi} = \frac{1}{n-1} \sum_{i=1}^{n} (OPD_c(n) - \mu_{OPD_c})^2 \ ,$$

    where $\mu_{OPD}$ is the mean of the OPD of the cell.

    b. **Phase kurtosis**: This parameter measures whether the cell OPD distribution is peaked or flat.

    $$(12)\ Kurtosis_{\varphi} = \sum_{i=1}^{n} \frac{(OPD_c(n) - \mu_{OPD_c})^4}{\sigma_{\varphi}^4}.$$

    c. **Phase skewness**: This parameter measures the lack of symmetry of the cell OPD values from the mean value.

    $$(13)\ Skewness_{\varphi} = \sum_{i=1}^{n} \frac{(OPD_c(n) - \mu_{OPD_c})^3}{\sigma_{\varphi}^3}.$$

11. **Energy:** This parameter characterizes the cell texture (70).

$$(14)\quad E = \sum_{i=1}^{n} OPD_c(n)^2 \ .$$

It is important to mention that the OPD values have coupling between the cellular refractive index and the physical thickness (see Eq. (1)), and the entire analysis was performed directly on the OPD value without decoupling these parameters to allow single exposure mode per each instance of the sample, which is suitable for acquiring cells during fast flow. Note also, that all 15 parameters presented above are based on the quantitative OPD map and thus cannot be calculated based on simple bright-field microscopy or fluorescent microcopy (recording the intensity of light), which is the basis of conventional flow cytometry, or based on non-quantitative phase techniques such Zernike's phase contrast microscopy and differential interference contrast (DIC) microscopy.

Statistical analysis

In total, we acquired 106, 97, 71, 102, 118, and 163 OPD maps for cell lines Hs 895.Sk, Hs 895.T, WM-115, WM-266-4, SW-480, and SW-620, respectively. To evaluate statistical difference amongst the cell groups, for each cell line pair and for each of the calculated parameters, we used the two-sample t-test for the p values of the data shown in Table 1. In addition, we also implemented the Mann-Whitney test, which yielded similar or even greater statistical difference between the groups.

Machine learning

Following the OPD-map-based feature extraction, our objective was to obtain classification decisions. For this goal, we used state-of-the-art machine learning techniques to perform three separate classifications for each pair of cell lines: Hs 895.Sk versus Hs 895.T, WM 115 versus WM 266-4, and SW 480 versus SW 620.

In general, a classification solution comprises of three main stages: (a) selection of features for the specific task, (b) a dimensionality reduction (and noise removal), and (c) the final classification using a selected classifier. In this work, we used the proposed analysis of the descriptors to select the features.

As for stage (a) of feature selection, for each cell OPD map, a set of 13 features which were found to be statistically discriminative according to their p values (mean, median, projected surface area, phase volume, dry mass, dry mass average density, energy, surface area, phase surface area to volume ratio, phase surface area to dry mass ratio, projected area to volume ratio, sphericity and phase variance), were concatenated into a feature vector of size 13. For each classification task, a feature matrix of size (M+N)×13 was built, where M and N are the numbers of cells in the matching cell line pair.

For stage (b) of dimensionality reduction, a standard Principal Component Analysis (PCA) (71) was performed on the feature matrix. The resulting PCA feature matrix was column normalized to a mean of 0 and a standard deviation of 1. A set of the most informative components can be selected following the PCA stage (71). In this work, after trying other options, we eventually selected the first six components to serve as the input representation.

Next, in stage (c) of classification, since we know all of the cells labels, a supervised learning method was selected. A state-of-the-art Support Vector Machine (SVM) classifier was used for this task. The SVM is widely used for pattern classification problems (72). A Leave-One-Out SVM classification was performed using LibSVM (73). The classification performance was evaluated using the area under curve (AUC) of the receiver operating characteristic (ROC) curve.

**Results**

We used the low-coherence IPM system presented in Fig. 1 to acquire OPD maps of the six cells lines described above, with the goal of comparing healthy to cancer cells (Hs 895.Sk [skin] vs. Hs 895.T [melanoma]), and primary cancer to metastatic cancer cells (WM-115 [melanoma] vs. WM-266-4 [metastatic melanoma], and SW-480 [adenocarcinoma colon] vs. SW-620 [metastatic adenocarcinoma colon]). Note that the cells were unattached and therefore had mostly round projected areas. The ability to analyze unattached isolated cells is important for flow cytometry via high-throughput quantitative imaging of cells during flow. Visualization 1 (Fig. 2) presents the OPD map of SW 480 cells flowing in a microfluidics channel (IBIDI, SN. 80666, 1 mm width, 17 mm length, 0.1 mm height). In this situation, when the isolated cells are round and unattached, most of the cells look alike and subjective pathological examination cannot be performed. Indeed, Fig. 3 shows one representative OPD map from each of the cell line groups, demonstrating that even when using quantitative phase imaging, a bare eye cannot see significant differences between each pair of cell lines. This is solved by the automatic machine-learning method analyzing the cell topological OPD maps directly. On the other hand, non-quantitative imaging methods, such as label-free bright-field, Zernike's phase contrast or DIC microscopies, which do not have access to these cell topological OPD maps, do not allow for quantitative automatic identification.

Prior to the analysis of these OPD maps, we applied the segmentation image processing procedure, described in the Methods section, to track the cell edges. Next, we applied Eqs. (1-14) to the cell OPD area selected by the segmentation process to calculate the following parameters: mean, median, projected surface area, dry mass, dry mass average density, phase volume, phase surface area, phase surface area to volume ratio, phase surface area to dry mass ratio, projected area to volume ratio, sphericity, phase variance, kurtosis, skewness and energy.

The corresponding average and standard deviation values for each parameter for all cell lines are summarized in Table 1. As can be seen from this table, 13 out of 15 OPD-based parameters were statistically significant. Figures 4-6 show the histograms of each of 12 parameters for which the two groups of cells compared were statistically significant (excluding the projected surface area parameter, to spare space, that was also statistically significant with p-values of <0.0005). Each subfigure shows the histograms of a pair of cell lines, where each cell line is shown in a different color (red or purple). All these parameters are statistically significant based on p-values of <0.05, <0.005 and even <0.0005. These results demonstrate the parameters ability to statistically discriminate between each cell line in each pair, even if the initial OPD map of the cells look similar.

The 13 successful parameters were used as an input for PCA/SVM analysis (see machine learning details in the Method section) to extract the best combination of these parameters, which is useful for classification between each cell line pair. The ROC curves obtained by using PCA followed by an SVM classification for each pair of cells are presented in Fig. 7. The best results of the classification were obtained for six-first principle components and for linear SVM kernel for all cell line pairs. As seen in this figure, the area under the curve (AUC) in all three classification tasks is high and around 0.9.

Table 2 summarizes the results obtained using the SVM learning combined with PCA for each pair of cells, with the AUC and the matching working points (sensitivity/specificity) chosen on the ROC curves. Note that for the definition of sensitivity and specificity, the Hs 895.Sk, SW 480 and WM 115 cell lines were defined as 'negatives', and the Hs 895.T, SW 620 and WM 266-4 cell lines were defined as 'positives'.

These results demonstrate an automatic algorithm with an ability to classify cells in different cancer stages using the OPD-map-based parameters. We achieved high classification rates and AUC values for unattached cells, whereby no morphological differences can be observed in the phase maps by the naked eye (see Fig. 3). The AUC values correspond with the separation between the groups presented in the histograms (Figs. 4-6), whereby better classification with higher percentages corresponds with lower p-values between the groups in the histograms. Thus, for the WM pair (Fig. 5), we can observe better separation than the Hs pair (Fig. 4) and the SW pair (Fig. 6).

Significantly, since the best PCA/SVM parameters were achieved for all cell line pair classification tasks (six-best principle components and linear SVM kernel were finally chosen for all cases), it can be assumed that a global machine for cancer grading can be developed.

**Discussion and conclusions**

Early detection of cancer can prevent tedious and painful treatment process, may prevent recurrence, and improve survival. To identify cancer, pathologists typically use fresh samples that are frozen and sectioned, or fixed samples that are dehydrated, embedded in paraffin, sectioned and stained with dyes and/or antibodies to specific tumor antigens. Our work describes the first steps in the development of optical signatures that can distinguish normal cells from cancer in situ and cancer in situ from metastatic forms without fixation, sectioning or any type of labeling. Thus, the scores can be performed rapidly and automatically and are not only based on simple qualitative cell parameters based on 2-D imaging (such as cell size and general shape), but rather use the cellular optical thickness dimension as well. This yields new parameters based on the cells topological map that have not been available to the pathologists before, with which automatic cell identification can be performed.

The challenge with flow cytometry for cancer diagnosis is the number of cells that are needed, which in turn is dependent on the brightness of the marker. In general, 1000 to 1,000,000 cells are needed from a dispersed tumor biopsy depending on the proportion of cancer cells to non-cancer cells in the sample and tumor heterogeneity (74). There is no universal marker that can be used to detect all cancers. Flow cytometry is used to detect leukemia, other bone marrow-associated cancers and to determine success of stem cell transplantations, but certainly cannot be used to detect circulating tumor cells where greater than or equal to five metastatic breast cancer cells are found in 7.5 mL peripheral blood (75). On the other hand, our method can distinguish between normal and cancerous cells and primary cancer from metastatic cancer cells, with as few as tens of cells, without using fixation, embedding and tumor markers.

To address the clinical need for cancer prognosis, we identified OPD-based diagnostic signatures for live and unattached cells in different stages of oncogenesis. We proposed using spatial morphological and texture parameters, which are based on the cell OPD maps, to compare tumor-derived cancer cell lines of different cancer levels and cell lines derived from a healthy tissue of the same individuals. We showed the feasibility to distinguish between different cell conditions in a label free manner and we have quantified differences among microscopically similar looking cells with statistical significance. We demonstrated that diagnostic optical signatures can be derived from comparisons between single cells, normal fibroblasts derived from living biopsy tissue compared to fibroblasts isolated from melanoma or from comparison of fibroblasts derived from melanoma to cells derived from a melanoma lymph metastasis. Thus, we showed that these quantitative phase based parameters are able to distinguish cancerous cells from the healthy cells and metastatic cancer cells from primary cancer cells with high accuracy.

Specifically, to compare healthy and cancerous cells, normal human skin fibroblasts isolated from a 48 year old Caucasian female (Hs 895.Sk) were compared to fibroblasts isolated a melanoma tumor from the skin of the same individual (Hs.895.T), with classification results of 81% sensitivity and 83% specificity. To compare primary cancer cells and metastatic cancer cells, human fibroblasts isolated from a melanoma tumor *in situ* from a 58 year old female (WM-115) were compared to fibroblasts isolated from a lymph metastasis in the same individual (WM-266-4), with classification results of 93% sensitivity and 99% specificity. Additionally, cells from colorectal adenocarcinoma in situ (SW-480) were compared to cells from colorectal adenocarcinoma lymph metastasis from the same individual (SW-620), with classification results of 82% sensitivity and 81% specificity.

Tumor microenvironment may differ between melanoma *in situ* and distant sites, contributing to reorganization of cellular morphology and interaction with quantitative phase. Indeed, metastatic forms of different cancers share morphological similarities leading to the idea that many cancers evolve into a more common metastatic state that is stratified by heterogeneity in tumor microenvironment, which differs depending on the site of metastasis, interaction with the immune system, and chemotherapeutic resistance (76).

Cells in our study were released from the substratum in adherent culture by limited digestion with trypsin resulting in spherical cells in solution, to model the geometry of circulating tumor cells. We compared the OPD profiles of trypsinized spherical normal cells to trypsinized spherical cancer cells and of trypsinized spherical primary cancer cells to trypsinized spherical metastatic cancer cells. The likely contributor to the difference in refractive index is that the cytoskeleton is altered dramatically during oncogenesis. Future experiments will determine how trypsinized spherical cells compare to affinity purified circulating tumor cells. We predict that the circulating tumor cells will resemble trypsinized cells analyzed in our experiments because cells inside a tumor are essentially in 3D adherent culture.

The optical signatures that can be built using the OPD parameters are advantageous in that tumor cells can be imaged without time consuming labeling, sectioning, sequencing or other methods in use today to identify tumor cells by genomic changes.

Thirteen of the parameters defined, which include the OPD mean, median, projected surface area, phase volume, dry mass, dry mass average density, surface area, phase surface area to volume ratio, phase surface area to dry mass ratio, projected area to volume ratio, sphericity, phase variance and energy, showed statistical significance with low p-values for all cell line pairs. Significantly, despite the wide distributions in the features of the various groups, as shown in the histograms in Figs. 4-6, which could be a result from differences in cell cycle length and non-synchronization across the population; low p-values were still obtained between the groups, so that these groups are statistically different. No statistical significance was observed for kurtosis and skewness, which might be explained by the fact that the experiments were done on floating, uniformly spherical cells, so the measures of peaks, flatness or symmetry were less applicable.

It can be seen in Figs. 4-6 that the parameter values decreased with cancer progression (healthy > cancer > metastatic) except for the projected area to volume ratio and sphericity (where the mathematical relation is opposite) that increased with cancer progression. In any case, as can be seen from these results, a similar trend of progression is retained in the average parameters values. We suppose that the observed trends will be kept for other similar cell line pairs. However, since our experiments are based on three cell line pairs, further research is needed to prove this hypothesis. The greater differences in the calculated parameters between the WM cell line cells than between isogenic SW cell line cells might be explained by the fact that that the WM fibroblasts isolated from *in situ* melanoma compared to the WM fibroblasts isolated from metastatic melanoma are likely to be more similar (both flat well-spread) than the SW epithelial cells isolated from colorectal adenocarcinoma, which are more cuboidal *in situ*, compared to SW metastatic cells, which are more spread and more motile.

IPM of unattached cells in flow can be performed faster than video processing rate on a regular computer, including 2-D phase unwrapping (56,77) and much faster if using the graphics processing unit (GPU) of the computer (78). Our method can be combined in the clinics using compact modules that can be connected in the output of a conventional microscope (79,80). By combining IPM with real-time fast processing algorithms (56,77,78) and automatic cell detection and machine learning algorithm for evaluating cell condition as presented in our work, quantitative phase microscopy has potential to become a powerful clinical screening tool for cancer diagnosis and might allow pathologists to inexpensively grade circulating tumor cells in liquid biopsies in real time. Thus, together with the analysis tools presented here for cancer monitoring, we believe that in the future IPM can be useful for detecting and monitoring cancer from body fluids in flow cytometry, for routine cell analysis or for the investigation and detection of pathological conditions in a semi-automatic way.

In this work, we used state-of-the-art machine-learning algorithms based on SVM to demonstrate the feasibility of our approach for automatic classification of cancer cells based on IPM data. However, note that there is a great variety of other machine leaning algorithms. One of the possible extensions is neural network related algorithms for deep learning, which require a greater number of samples.

To summarize, the proposed quantitative imaging technique shows preliminary clinical potential for automatic cell flow cytometry. This technique does not require cell staining as conventional flow cytometry, and since it uses the quantitative phase profile it has access to parameters which are not available using bright-field microscopy or other non-quantitative phase imaging techniques. Still, future studies are needed to confirm if the unique quantitative phase signatures developed here can distinguish normal from circulating cancer cells isolated from liquid biopsies, and determine the diagnostic and prognostic value of quantitative phase signatures in determining tumor grade and metastatic potential.

**Acknowledgments**

This research was supported by Horizon 2020 European Research Council (ERC) Grant No. 678316. We thank Ruth Gottlieb for significant biological support and advising, Dr. Ksawery Kalinowski for useful technical discussions, and Dr. Itay Barnea for useful biological discussions.

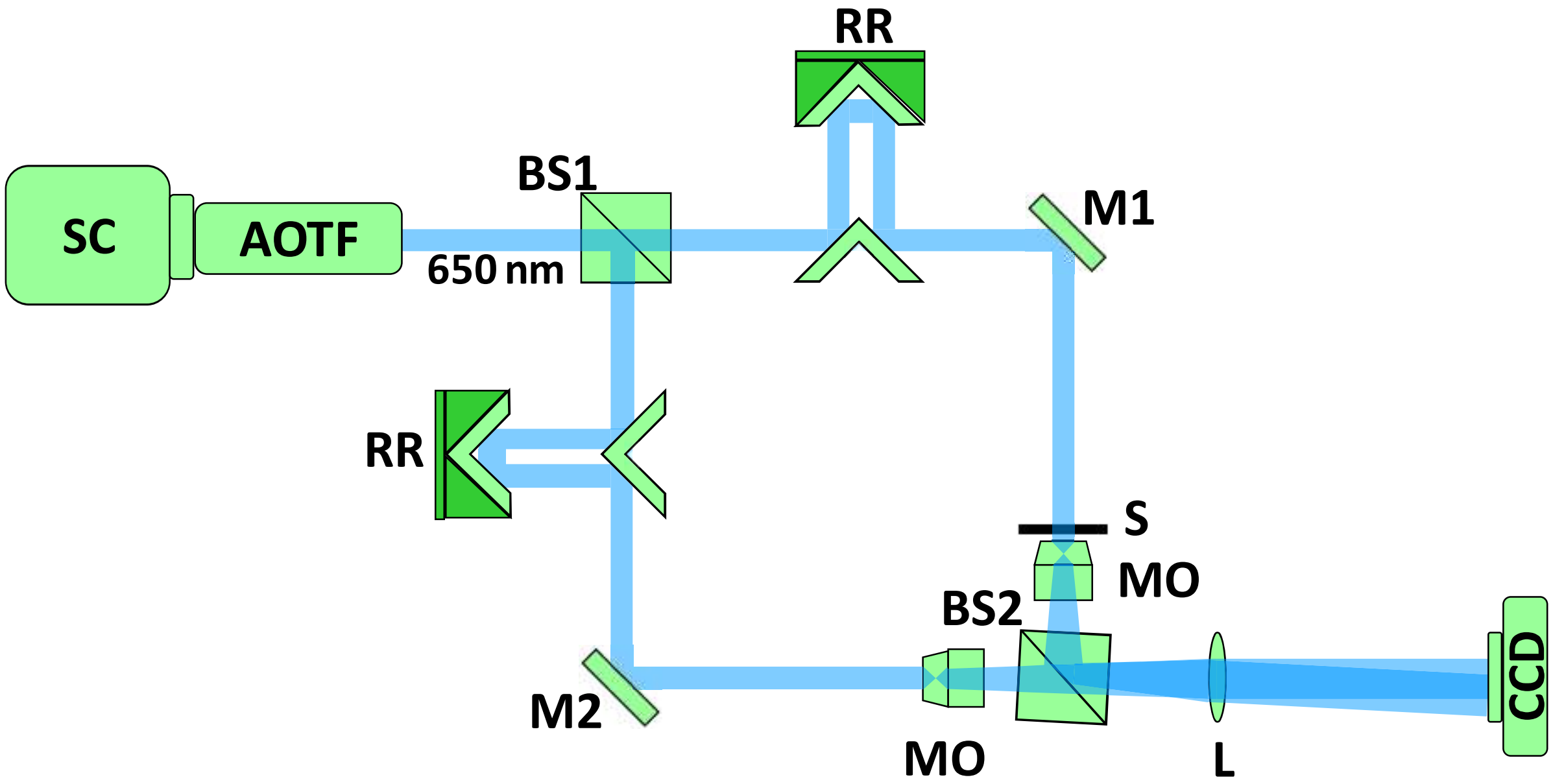

**Figure. 1.** Low-coherence IPM used in the experiments SC, Super continuum laser source. AOTF, Acousto optical tunable filter. BS1, BS2, Beam splitters. M1, M2, mirrors. RR, retroreflector. S, Sample. MO, Microscope objective. L, tube lens. CCD, Digital camera.

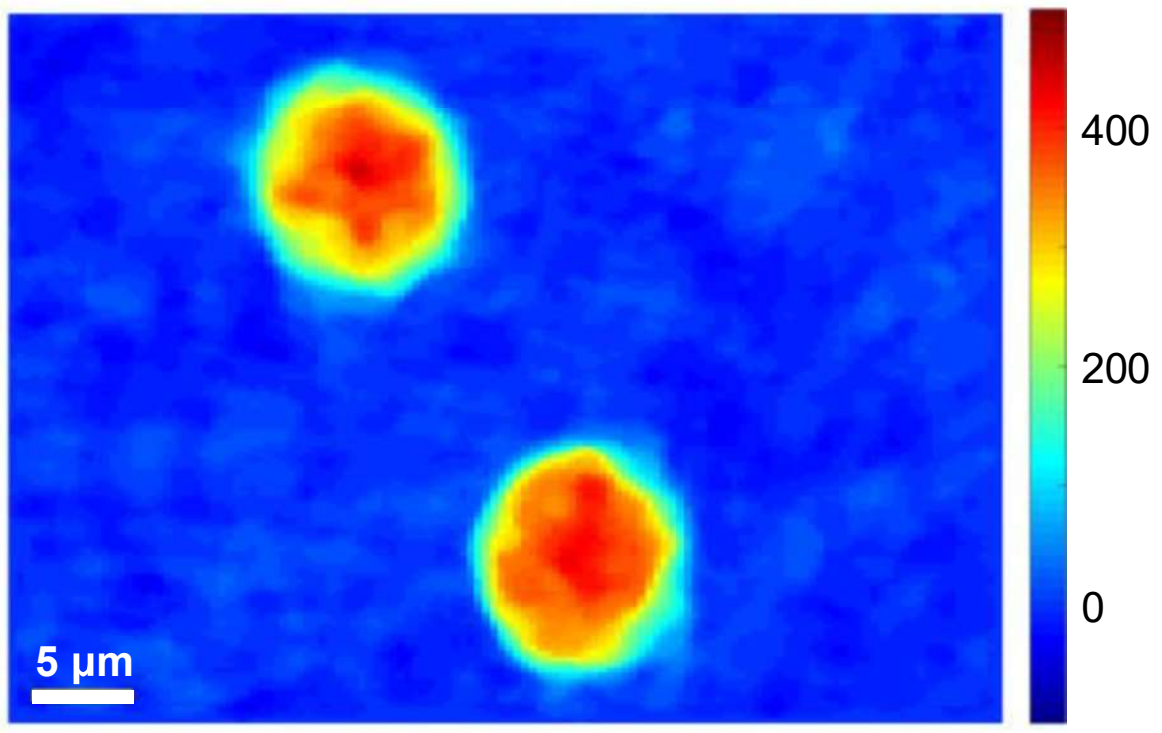

**Figure. 2.** One snapshot from a video (see Visualization 1) presenting quantitative phase microscopy of live unlabeled cancer cells (SW 480 cell line) flowing in a microfluidic channel. Colorbar represents OPD values in nm.

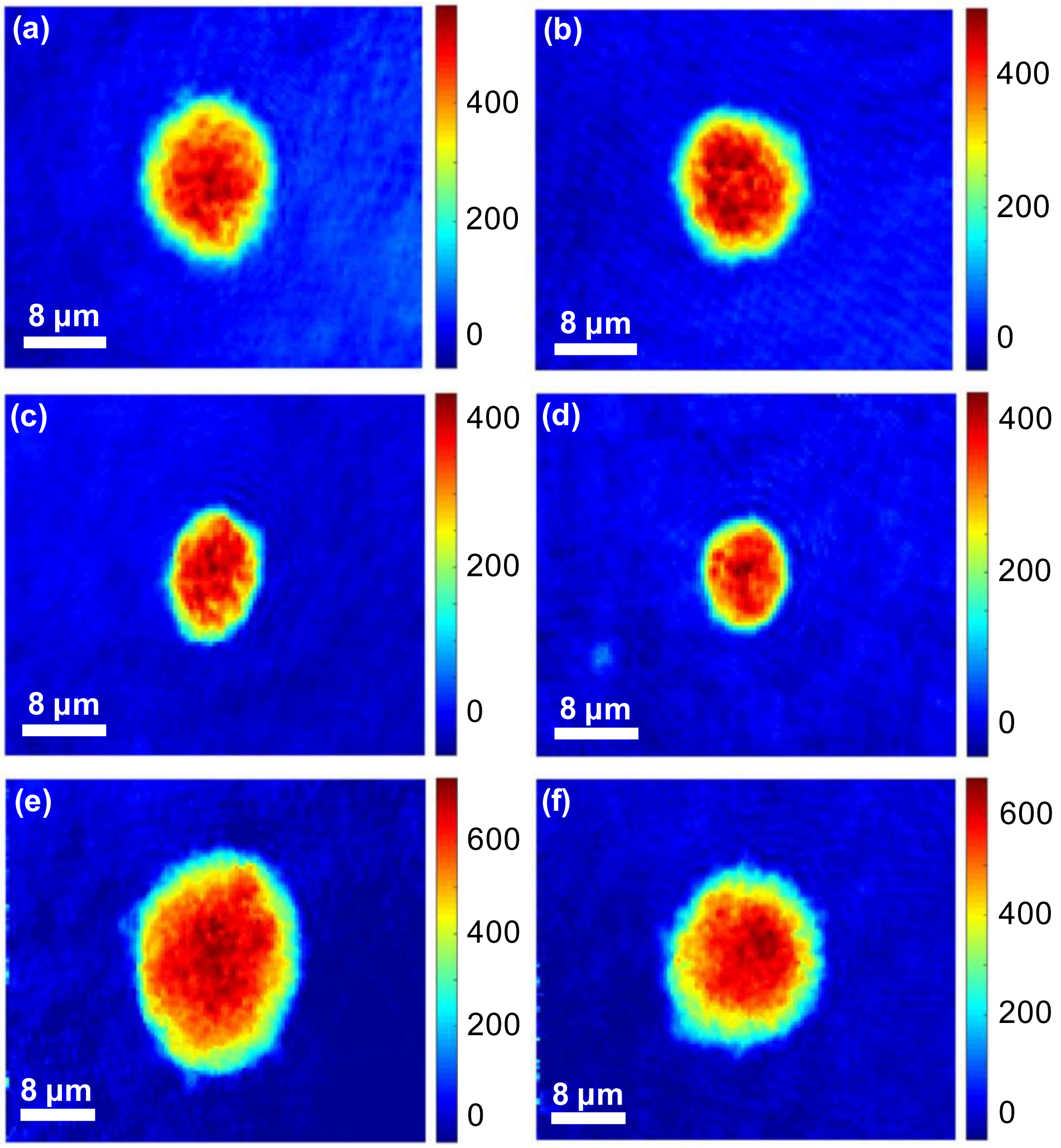

**Figure. 3.** Quantitative phase microscopy of unattached cancer cells, demonstrating that most cells are round and look alike in this situation. (a) Hs 895.Sk. (b) Hs 895.T. (c) SW480. (d) SW620. (e) WM 115. (f) WM 266-4. Colorbars represent OPD values in nm.

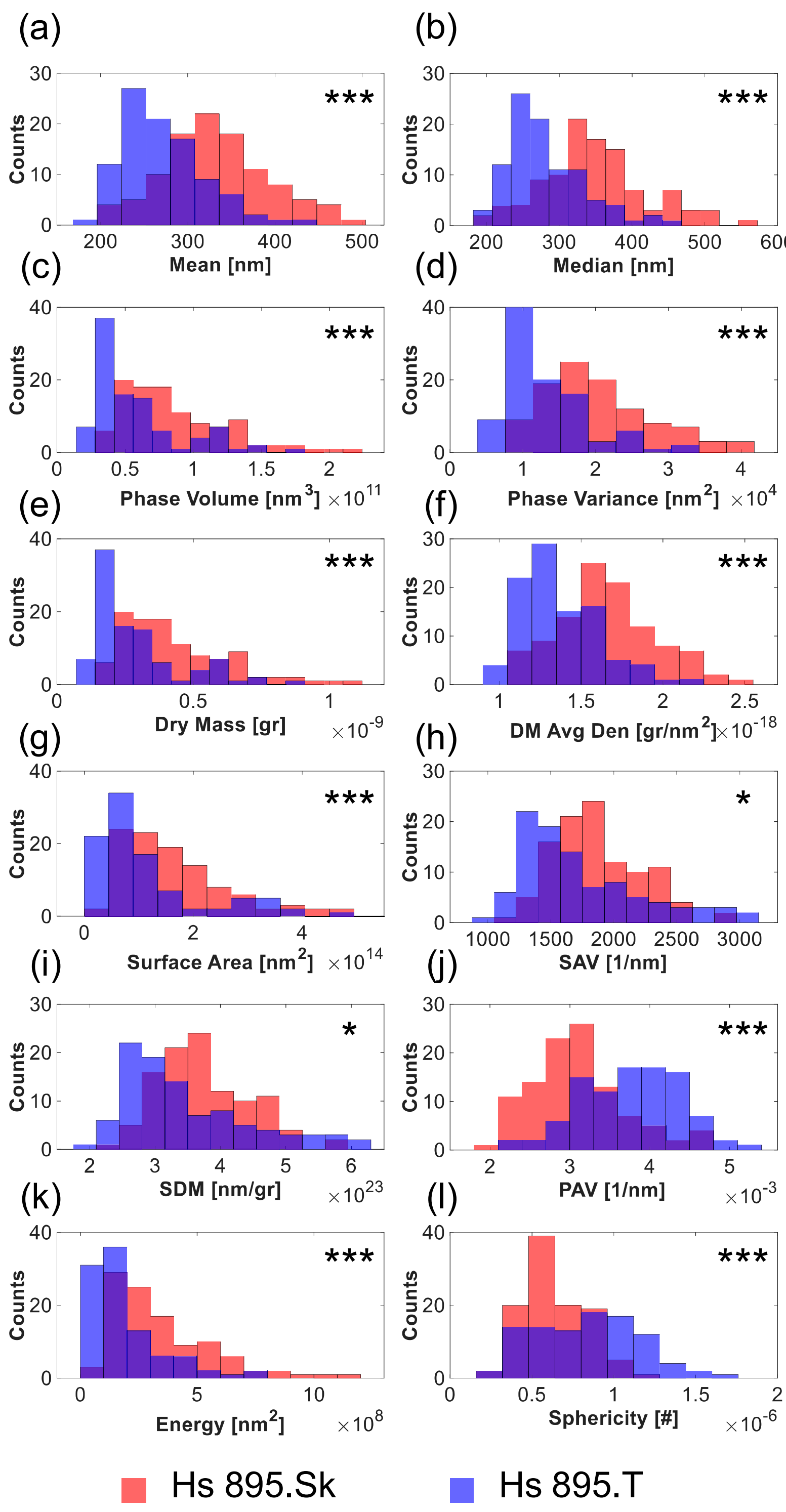

**Figure. 4.** Histograms of the parameters based the OPD maps for normal skin cells Hs 895.Sk (red) via melanoma skin cells Hs 895.T parameters (blue). (a) Mean. (b) Median. (c) Phase Volume. (d) Phase Variance. (e) Dry Mass. (f) Dry Mass Average Density. (g) Surface Area. (h) Phase Surface Area to Volume Ratio. (i) Phase Surface Area to Dry Mass Ratio. (j) Projected area to volume ratio. (k) Energy. (l) Sphericity. * denotes p-value<0.05, *** denotes p-value<0.0005.

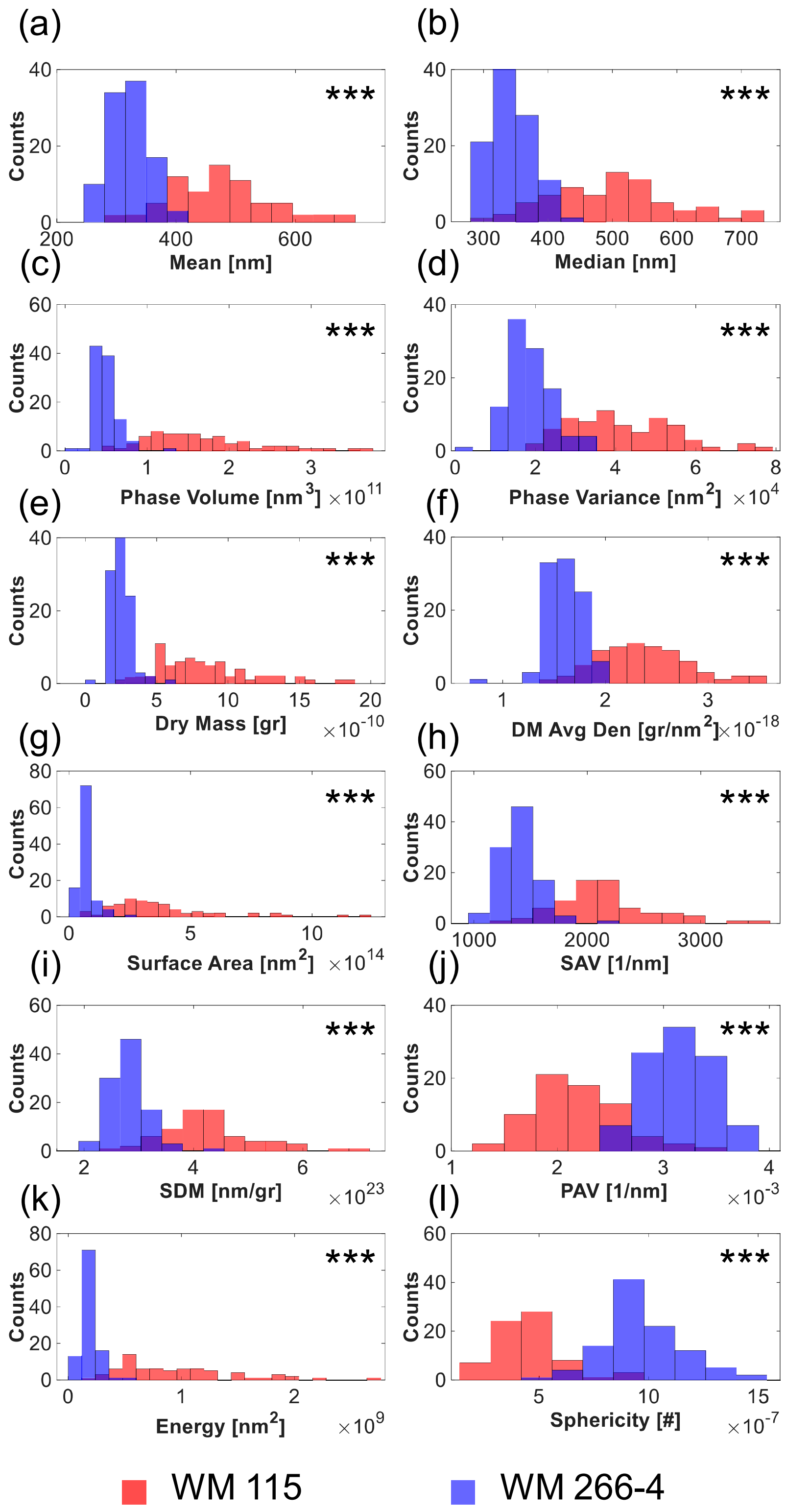

**Figure. 5.** Histograms of the parameters based the OPD maps for melanoma skin cells WM-115 (red) via metastatic melanoma skin cells WM-266-4 (blue). (a) Mean. (b) Median. (c) Phase Volume. (d) Phase Variance. (e) Dry Mass. (f) Dry Mass Average Density. (g) Surface Area. (h) Phase Surface Area to Volume Ratio. (i) Phase Surface Area to Dry Mass Ratio. (j) Projected area to volume ratio. (k) Energy. (l) Sphericity. *** denotes p-value<0.0005.

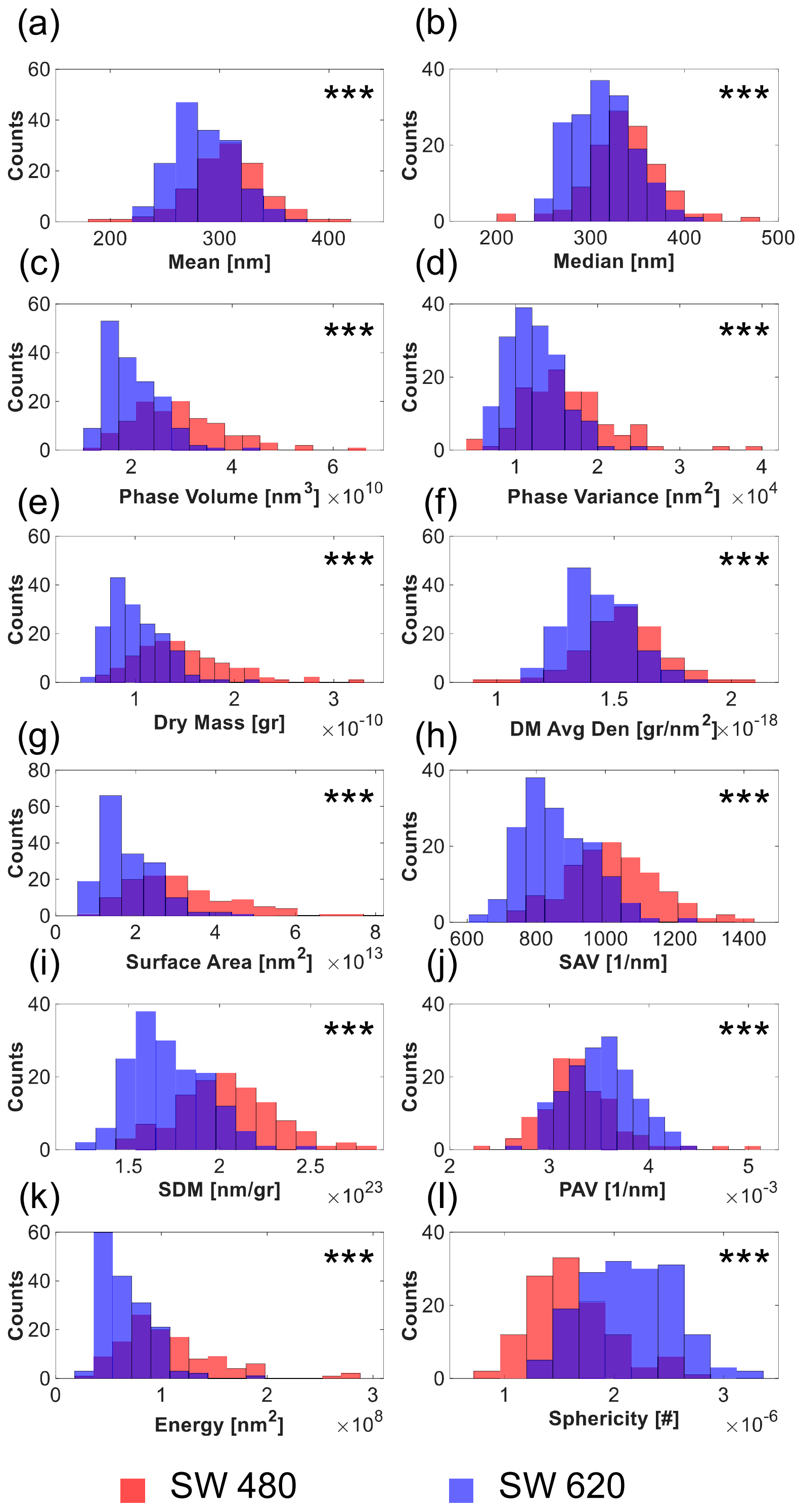

**Figure. 6.** Histograms of the parameters based the OPD maps for colorectal adenocarcinoma colon cells SW-480 (red) via metastatic from lymph node of colorectal adenocarcinoma cells SW-620 (blue). (a) Mean. (b) Median. (c) Phase Volume. (d) Phase Variance. (e) Dry Mass. (f) Dry Mass Average Density. (g) Surface Area. (h) Phase Surface Area to Volume Ratio. (i) Phase Surface Area to Dry Mass Ratio. (j) Projected area to volume ratio. (k) Energy. (l) Sphericity. *** denotes p-value<0.0005.

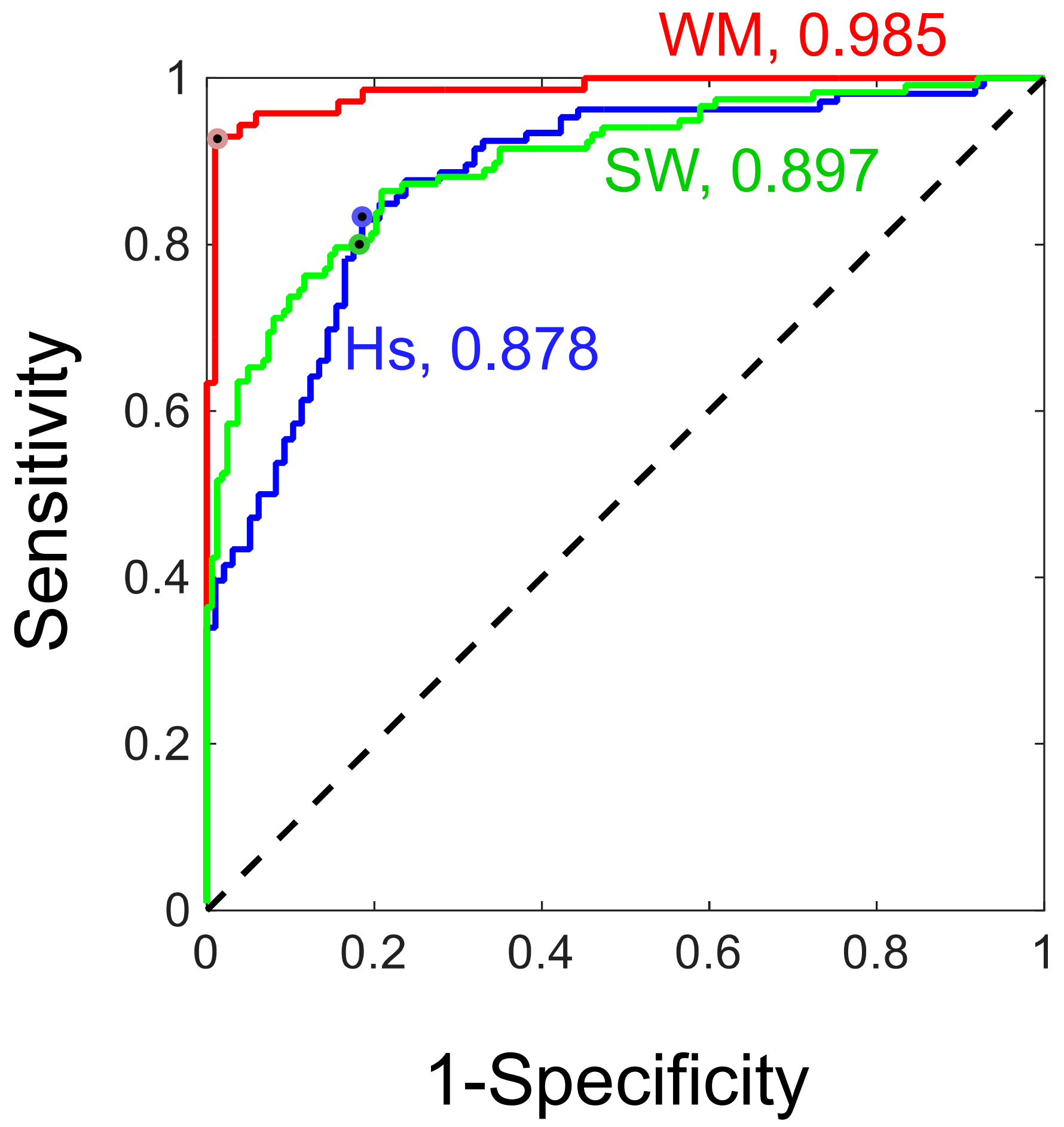

**Figure 7.** ROC curves of true-positive rate (sensitivity) vs. false-positive rate (one minus specificity), as obtained using SVM learning combined with PCA for the: Hs cell lines (blue curve), WM cell lines (red curve), SW cell lines (green curve). Each graph represents another cell line pair (Hs, WM, or SW), with the resulting AUC written after the cell line pair name. The small circles denote the working points of the classification tasks.

**Table 1.** Averages and standard deviations for all the parameters calculated for all the cell lines. In each cell, the first and second values correspond to the average and to the standard deviation values, respectively. * denotes p-value<0.05, ** denotes p-value<0.005, *** denotes p-value<0.0005. - denotes p-value>0.05.

| | Hs 895.Sk | Hs 895.T | | WM 115 | WM 266-4 | | SW 480 | SW 620 | |
|---|---|---|---|---|---|---|---|---|---|
| Mean [nm] | 333.09<br>60.99 | 272.40<br>48.45 | *** | 474.10<br>86.58 | 319.90<br>35.70 | *** | 306.45<br>34.59 | 286.04<br>27.82 | *** |
| Median [nm] | 348.88<br>72.37 | 282.53<br>52.90 | *** | 502.78<br>93.88 | 342.90<br>39.04 | *** | 335.89<br>40.07 | 312.34<br>32.95 | *** |
| Projected surface area [$10^{6}\cdot nm^{2}$] | 251.87<br>78.94 | 202.99<br>84.96 | *** | 342.24<br>106.06 | 153.14<br>34.62 | *** | 95.60<br>21.48 | 70.4<br>13.53 | *** |
| Phase volume [$10^{9}\cdot nm^{3}$] | 86.34<br>38.76 | 57.67<br>33.01 | *** | 165.93<br>69.02 | 49.75<br>14.87 | *** | 29.74<br>9.29 | 20.34<br>5.35 | *** |
| Dry mass [$10^{-11}\cdot gr$] | 43.17<br>19.38 | 28.84<br>16.51 | *** | 82.96<br>34.51 | 24.87<br>7.44 | *** | 14.87<br>4.64 | 10.17<br>2.68 | *** |
| Dry mass average density [$10^{-19}\cdot gr/nm^{2}$] | 16.65<br>3.05 | 13.62<br>2.42 | *** | 23.70<br>4.33 | 15.99<br>1.79 | *** | 15.32<br>1.73 | 14.30<br>1.39 | *** |
| Surface area [$10^{12}\cdot nm^{2}$] | 172.16<br>109.01 | 114.37<br>100.90 | *** | 376.22<br>225.56 | 71.57<br>31.83 | *** | 31.36<br>13.90 | 17.96<br>6.93 | *** |
| SAV [$10^{2}\cdot 1/nm$] | 18.71<br>3.53 | 17.37<br>4.92 | * | 21.46<br>4.19 | 13.87<br>2.08 | *** | 10.17<br>1.34 | 8.59<br>1.04 | *** |
| SDM [$10^{22}\cdot nm/gr$] | 37.43<br>7.06 | 34.75<br>9.84 | * | 42.92<br>8.38 | 27.75<br>4.16 | *** | 20.35<br>2.68 | 17.19<br>2.08 | *** |
| PAV [$10^{-4}\cdot 1/nm$] | 31.05<br>5.84 | 37.77<br>6.17 | *** | 21.80<br>4.06 | 31.74<br>4.64 | *** | 33.07<br>4.01 | 35.29<br>3.40 | *** |
| Sphericity [$10^{-7}\cdot \#$] | 6.42<br>1.95 | 8.45<br>3.16 | *** | 4.50<br>1.44 | 11.26<br>15.59 | *** | 16.13<br>3.77 | 21.39<br>4.12 | *** |
| Phase variance [$10^{3}\cdot nm^{2}$] | 20.58<br>7.46 | 13.22<br>5.69 | *** | 41.38<br>12.84 | 18.85<br>5.31 | *** | 16.06<br>5.34 | 12.34<br>3.29 | *** |
| Phase kurtosis [$10^{-6}\cdot 1/nm^{4}$] | 15.02<br>10.42 | 29.29<br>18.86 | *** | 4.85<br>3.10 | 11.41<br>17.57 | ** | 10.24<br>11.09 | 11.79<br>6.50 | – |
| Phase skewness [$10^{-5}\cdot 1/nm^{3}$] | -21.57<br>23.91 | -26.75<br>21.40 | – | -15.04<br>8.60 | -21.86<br>21.33 | * | -25.29<br>16.67 | -28.89<br>14.19 | – |
| Energy [$10^{7}\cdot nm^{2}$] | 34.97<br>21.54 | 19.30<br>15.01 | *** | 94.82<br>52.08 | 18.72<br>7.40 | *** | 10.64<br>4.60 | 6.61<br>2.38 | *** |

**Table 2.** Machine learning results for classifications between the groups (PCA-SVM analysis).

| | Sensitivity | Specificity | AUC |
|---|---|---|---|
| Hs 895.Sk<br>Hs 895.T | 81% | 83% | 0.878 |
| WM 115<br>WM 266-4 | 93% | 99% | 0.985 |
| SW 480<br>SW 620 | 82% | 81% | 0.897 |